\documentclass[conference]{IEEEtran}
\IEEEoverridecommandlockouts

\usepackage{cite}
\usepackage{amsmath,amssymb,amsfonts}
\usepackage{algorithmic}
\usepackage{graphicx}
\usepackage{textcomp}
\usepackage{url}
\usepackage{graphicx}
\usepackage{bm}
\usepackage{multirow}
\usepackage{booktabs}
\usepackage{makecell}
\usepackage{bbding}
\usepackage{amssymb}
\usepackage{pifont}
\usepackage{wasysym}
\usepackage{utfsym}
\usepackage{fontawesome}
\usepackage{balance}
\usepackage{float}
\usepackage{amsmath}
\usepackage{amsfonts}
\usepackage{bm}
\usepackage{color}
\usepackage{hyperref}
\usepackage{algorithmic}
\usepackage{algorithm}
\usepackage{mathrsfs}
\usepackage{colortbl}
\hypersetup{
    colorlinks=true,
    linkcolor=black,
    filecolor=black,      
    urlcolor=black,
    citecolor=black,
}
\usepackage{hyperref}
\makeatletter
\def\UrlAlphabet{%
      \do\a\do\b\do\c\do\d\do\e\do\f\do\g\do\h\do\i\do\j%
      \do\k\do\l\do\m\do\n\do\o\do\p\do\q\do\r\do\s\do\t%
      \do\u\do\v\do\w\do\x\do\y\do\z\do\A\do\B\do\C\do\D%
      \do\E\do\F\do\G\do\H\do\I\do\J\do\K\do\L\do\M\do\N%
      \do\O\do\P\do\Q\do\R\do\S\do\T\do\U\do\V\do\W\do\X%
      \do\Y\do\Z}
\def\UrlDigits{\do\1\do\2\do\3\do\4\do\5\do\6\do\7\do\8\do\9\do\0}
\g@addto@macro{\UrlBreaks}{\UrlOrds}
\g@addto@macro{\UrlBreaks}{\UrlAlphabet}
\g@addto@macro{\UrlBreaks}{\UrlDigits}
\makeatother
\usepackage{threeparttable}
\usepackage{tabu}
\def\BibTeX{{\rm B\kern-.05em{\sc i\kern-.025em b}\kern-.08em
    T\kern-.1667em\lower.7ex\hbox{E}\kern-.125emX}}
\begin{document}

\title{Leveraging LLM and Text-Queried Separation for Noise-Robust Sound Event Detection
}

\author{\IEEEauthorblockN{
Han Yin$^{1}$, Yang Xiao$^2$, Jisheng Bai$^1$, Rohan Kumar Das$^2$}
\IEEEauthorblockA{\textit{$^1$School of Marine Science and Technology, Northwestern Polytechnical University, Xi’an, China}}
\IEEEauthorblockA{\textit{$^{2}$Fortemedia Singapore, Singapore}}
}


\maketitle

\begin{abstract}

Sound Event Detection (SED) is challenging in noisy environments where overlapping sounds obscure target events. Language-queried audio source separation (LASS) aims to isolate the target sound events from a noisy clip. However, this approach can fail when the exact target sound is unknown, particularly in noisy test sets, leading to reduced performance. To address this issue, we leverage the capabilities of large language models (LLMs) to analyze and summarize acoustic data. By using LLMs to identify and select specific noise types, we implement a noise augmentation method for noise-robust fine-tuning. The fine-tuned model is applied to predict clip-wise event predictions as text queries for the LASS model. Our studies demonstrate that the proposed method improves SED performance in noisy environments. This work represents an early application of LLMs in noise-robust SED and suggests a promising direction for handling overlapping events in SED. Codes and pretrained models are available at \url{https://github.com/apple-yinhan/Noise-robust-SED}.
\end{abstract}


\section{Introduction}

Sound event detection (SED) \cite{sed} involves identifying sound event classes and corresponding timestamps, which has been widely applied in various applications such as smart city \cite{smarthome,smarthome2} and medical monitoring \cite{medical}. Recent deep learning advancements have significantly enhanced SED, supported by diverse datasets designed for specific scenarios. One such widely used dataset is domestic environment sound event detection (DESED)~\cite{sed2}, which is also considered for Task 4 of the detection and classification of acoustic scenes and events (DCASE)\footnote{\url{https://dcase.community/challenge2024/}\label{dcase2024}} challenge series. Convolutional recurrent neural network (CRNN)-based models \cite{crnn4sed,fmsg2,fmsg_dcase2024,jiangyidi} have performed well on the DESED dataset in recent years. However, these studies typically consider testing data that closely matches the training data conditions, limiting the models' ability to generalize to real-world conditions~\cite{xiao2024mixstyle}. In noisy environments, audio may include sound events that were not be present during training, which can act as background noise, overlap with target sounds, and make detection more difficult, leading to reduced performance. As a result, developing noise-robust SED system remains a challenging problem. 

The pursuit of developing noise-robust SED has led to different methods for improving the performance in noisy conditions. One common approach is to fine-tune models using noisy data, which involves selecting noise types that represent real-world environments while being distinct from target sounds \cite{noise1,noise2,noise3,noise4,noise5}. However, artificial selection can be difficult due to bias and unnatural correlations. Recently, large language models (LLMs) like GPT-4 \cite{gpt4}, GPT-3 \cite{gpt3}, and Llama \cite{llama} have shown great capabilities in analyzing the semantic relationships between different sound types. An LLM-powered dataset referred to as WildDESED was proposed in~\cite{wilddesed}, where SED models are trained with a noisy training set that included the same noise events as in the test sets. However, training of SED models in this way, may not be very ideal for real-world applications, as {\it predicting the types of noise events during testing is often not feasible}.

To avoid the issue of predicting noise types during testing, we propose to use audio source separation to extract target sounds from noisy audio during testing. This method does not rely on knowing the specific noise events beforehand. Instead, it isolates the target sound directly from the audio mixture, making it adaptable to various noisy conditions. Language-queried audio source separation (LASS) models allow them to isolate audio signals that match the text query by pre-training on large audio-text paired datasets \cite{lass}. However, the text query is crucial, as it directly impacts the separation result. These models can only isolate sound events present in the audio. For example, if ``dog barking" is requested in a clip containing only speech, the output may be unreliable. In noisy environments, the actual sound events are unknown, and therefore it is difficult to generate accurate text queries for separation. To address this, we propose using LLMs to develop a noise-robust model that can generate reliable event predictions for text queries.

In this study, to address the challenge of noisy sound events in SED, we propose an LLM-based framework that integrates noise augmentation with the text-queried audio separation method for noise-robust SED. Our proposed approach first leverages LLMs to select appropriate noise types and fine-tune the SED model. The fine-tuned model then generates clip-wise event predictions, which serve as text queries for the LASS model during testing. Experimental studies in this work are conducted on DESED and WildDESED datasets to demonstrate the effectiveness of the proposed method to develop noise-robust SED system in challenging acoustic conditions. 
We introduce four different baseline methods for generating text queries to validate the effectiveness of our LLM-based approach. To the best of our knowledge {\it it is the first study to apply LASS models for noise-robust SED.} We have released the code to support future research.

\begin{figure*}[h]
\centering
\centerline{\includegraphics[width=\textwidth]{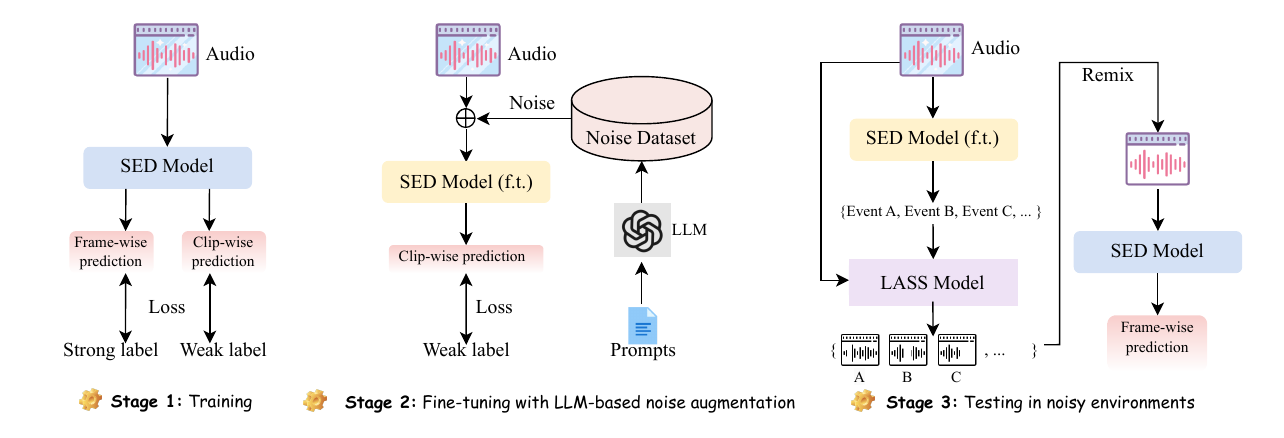}}
\vspace{-5mm}
\caption{The overview of LLM-based noise augmentation and text-queried separation for noise-robust sound event detection.}
\label{fig:overview}
\end{figure*}



\section{Proposed Method for Noise-robust SED}
As shown in Fig.~\ref{fig:overview}, the proposed noise-robust SED method can be decomposed into three stages: training, fine-tuning, and testing. Details are described as follows.

\subsection{Training}

During the training process, similar to conventional SED training approaches \cite{sed,crnn4sed}, we directly utilize an SED model to extract information from the input audio spectrogram  $\bm{X}\in\mathbb{R}^{T\times F}$, where $T$ and $F$ represent the number of frames and frequency bins, respectively.
\begin{equation}
    (\bm{\tilde{Y}}_\textrm{s}, \bm{\tilde{Y}}_\textrm{w}) = g_{\theta}(\bm{X})
\end{equation}
where $\bm{\tilde{Y}}_\textrm{s}\in\mathbb{R}^{T\times N}$ and $\bm{\tilde{Y}}_\textrm{w}\in\mathbb{R}^{N}$ are frame-wise and clip-wise predictions. 
$N$ is the number of target event classes. $g(\cdot)$ is the SED model, and $\theta$ is the training parameter.

Then, the binary cross entropy (BCE) loss between outputs and labels is calculated, which is formulated as:
\begin{equation}
    Loss_{\textrm{train}} = l_{\textrm{BCE}}(\bm{\tilde{Y}}_\textrm{s}, \bm{Y}_\textrm{s}) + l_{\textrm{BCE}}(\bm{\tilde{Y}}_\textrm{w}, \bm{Y}_\textrm{w})
\end{equation}
where $\bm{Y}_\textrm{s}\in\mathbb{R}^{T\times N}$ and $\bm{Y}_\textrm{w}\in\mathbb{R}^{N}$ are strong and weak labels.

\subsection{Fine-tuning with LLM-based noise augmentation}
Denote the training data of the training stage as $\mathcal{U}_\theta$, which mainly contains $N$ event classes.
However, in real-life scenarios, sound events may present a more complex and variable distribution, denoted as $\mathcal{U}_\alpha$, with $M$ event classes ($M \textgreater N$).
Because of the gap between $\mathcal{U}_\alpha$ and $\mathcal{U}_\theta$, the model trained on $\mathcal{U}_\theta$ (i.e. model $g_\theta$) may perform poorly on $\mathcal{U}_\alpha$.

To improve the performance of $g_\theta$ on $\mathcal{U}_\alpha$, we propose to use an LLM-based noise augmentation method to fine-tune $g_\theta$.
As shown in Fig.~\ref{fig:overview}, during fine-tuning, the LLM is used to select noise events from the noise candidate dataset.
We denote the distribution of the noise dataset as $\mathcal{U}_\gamma$, along with $K$ event classes ($K \textgreater N$).
Compared to $\mathcal{U}_\theta$, $\mathcal{U}_\gamma$ should present a more complex event distribution.
The selected noise signal is added to the input audio according to a specific signal-to-noise ratio (SNR).
An important question is: {\it how to set the input prompts of the LLM for selecting the noise?}

Inspired by WildDESED, we set prompts based on event classes of $\mathcal{U}_\gamma$ and $\mathcal{U}_\theta$.
Denote the weak annotation of the input audio as $\mathcal{A}=\{\bm{a}_i\}$, where $\bm{a}_i$ is the event that occurred in the clip and $i=1,2,3,...,K^{'}$. We first input event classes in $\mathcal{U}_\gamma$ to the LLM, then guide it by the following prompt: 

``{\it This is an audio clip recorded in a household environment. The following events present this clip: $\bm{a}_1, \bm{a}_2, ..., \bm{a}_{K^{'}}$.
Please help me select the event classes in $\mathcal{U}_\gamma$ that may occur in the audio clip.
Further, apply filtering to ensure that the selected classes do not have any classes similar to those already present in the audio clip.''}

The selected noise events are randomly added to the input audio signal according to a specific SNR.
To improve the fine-tuning stability and strengthen the model’s resilience to different noise levels, we apply a curriculum learning approach \cite{curriculum1,curriculum2,curriculum3,curriculum4,curriculum5}.
This approach gradually introduces the data complexity, beginning with clean samples and incrementally increasing the noise level. 
Throughout the fine-tuning process, we select appropriate noise events from $\mathcal{U}_\gamma$ by prompting the LLM, and these events are gradually added to the data distribution of $\mathcal{U}_\theta$. 
Through this procedure, the distribution of sound events in $\mathcal{U}_\theta$ gradually becomes more complex and closer to variable real-life data distribution of $\mathcal{U}_\alpha$.

\subsection{Testing in noisy environments}
When we test SED models in noisy environments, unpredictable noise sound events will alias with the target events, resulting in degraded detection performance.

Therefore, we utilize a pre-trained LASS model to extract target events from the noisy audio clip. The separated tracks are then remixed to produce relatively clean audio, reducing the influence of noisy events. As shown in Fig.~\ref{fig:overview}, the noisy audio spectrogram $\bm{X}^{'}\in\mathbb{R}^{T\times F}$ is first fed into the fine-tuned model $g_{\alpha}$, generating the clip-wise prediction $\bm{Y}_\textrm{w}^{'}\in\mathbb{R}^{N}$: 
\begin{equation}
    \bm{Y}_\textrm{w}^{'}=g_{\alpha}(\bm{X}^{'})
\end{equation}

A fixed threshold is applied to $\bm{Y}_w^{'}$ for indicating whether an event exists or not, which is set to $0.5$, same as DCASE 2024 Task 4 \cite{dcase2024T4}.
In this way, we can obtain the target events (e.g. ``speech'', ``dog'', ..., ``blender'') that are present in the input noisy audio clip.
These event texts are used as input queries for the pre-trained LASS model, denoted as $\mathcal{Q}=\{q_i| i=1,2,...,k\}$.
Combined with the noisy audio spectrogram, we can get the audio tracks corresponding to the text queries:
\begin{equation}
\begin{aligned}
\bm{s}_1 &= h(\bm{X}^{'}, q_1) \\
\bm{s}_2 &= h(\bm{X}^{'}, q_2) \\
...&\\
\bm{s}_k &= h(\bm{X}^{'}, q_k)
\end{aligned}    
\end{equation}
where $h(\cdot)$ is the pre-trained LASS model. The separated audio tracks are remixed as a mixture, as formulated in:
\begin{equation}
    \bm{s} = \frac{1}{k}\sum_{j=1}^{k}\bm{s}_j
\end{equation}
Compared to the input noisy audio, the remixed audio $\bm{s}$ contains fewer noise sound events, which is beneficial to reduce the interference of noise events on detecting target events.
The spectrogram of $\bm{s}$ is put into the SED model $g_{\theta}(\cdot)$ to generate the final frame-wise event prediction.

\section{Experimental Setup}
\subsection{Datasets}
\subsubsection{DESED} The DESED\footnote{\url{https://project.inria.fr/desed/description/}\label{desed}} dataset is designed to recognize 10 target sound event classes in domestic environments, which has been applied in DCASE challenges since 2020. 
For training, it provides a weakly-labeled set, a strongly-labeled set, and an unlabeled set. 
The weakly-labeled set contains 1,578 real recordings with weak annotations, capturing the presence of sound events without timestamps.
The unlabeled set comprises 14,412 real recordings.
The strongly-labeled set is composed of 10,000 synthetic recordings with strong annotations, representing exact temporal boundaries of sound events.
For validation, DESED provides 2,500 synthetic recordings with strong labels.
For testing, it contains 1,168 real recordings with strong annotations to evaluate performance.

\subsubsection{AudioSet-Strong} AudioSet \cite{audioset} consists of an expanding ontology of 632 audio event classes and a collection of 2,084,320 human-labeled 10-second sound clips drawn from YouTube videos.
AudioSet-Strong\footnote{\url{https://research.google.com/audioset/download_strong.html}\label{audioset}} is the strongly-labeled subset of AudioSet, along with 356 event classes.
It is important to note that not all event classes in AudioSet-Strong correspond to those in DESED. Following the DCASE guidelines, we use only the overlapping event classes from AudioSet-Strong for training and fine-tuning, resulting in a total of 3,470 audio clips. Meanwhile, the entire AudioSet-Strong dataset is used as the noise dataset for data augmentation in stage 2.

\subsubsection{WildDESED-Test} The WildDESED\footnote{\url{https://zenodo.org/records/13910598}\label{wilddesed}} dataset is an extension to DESED.
By leveraging LLMs, it adds sound events as background noise into the audio clips of DESED, aiming to bridge the gap between the controlled environment of existing datasets and the dynamic real-world domestic soundscapes, thus expanding the potential for noise-robust SED research in truly ``wild'' home scenarios.
The WildDESED dataset has four different SNR settings, from -5 dB to 10 dB. 
WildDESED-Test is the test set of it, and we use it for evaluation in stage 3. 

\subsection{Evaluation Metrics}
We evaluate systems using the threshold-independent polyphonic sound event detection scores (PSDS) \cite{psds1,psds2} in accordance with the DCASE 2024 Challenge Task 4 protocol, across two scenarios. 
Scenario 1 (P1) focuses on temporal localization precision, while scenario 2 (P2) prioritizes reducing confusion between different sound events.






\subsection{Implementation Details}
Following the DCASE 2024 Task 4 baseline, we use CRNN as the SED model, and apply the pre-trained model, BEATs \cite{beats}, to extract embeddings for better detection performance. 
We use AudioSep-DP \cite{audiosep-dp} as the pre-trained LASS model, which performs best on DCASE 2024 Task 9 objective single model benchmark\textsuperscript{\ref{dcase2024}}, composing of a ResUNet \cite{lass} with DPRNN \cite{dprnn} for separation and a CLAP \cite{clap} for text encoding.
We apply GPT-4 as the LLM for picking up noise data, which is an advanced language model that builds on the GPT-3 architecture but uses a larger amount of training data.
The SNR ranges from -10 dB to 10 dB during fine-tuning and batch size is set to 48. We use the Adam optimizer with a learning rate of 0.001 and an exponential warm-up scheduler is applied over the first 50 epochs of the total 200. 
To enhance training stability, we employ a mean teacher model \cite{mean-teacher}, setting the exponential moving average factor to 0.999.

\begin{table*}[t]
\centering
\caption{SED Results of different systems on DESED-Test and WildDESED-Test. The best metrics are in bold (excluding case \#2, which uses ground truth as text queries for audio source separation).}
\renewcommand\arraystretch{1.3}{
\setlength{\tabcolsep}{1mm}{
\begin{tabular}{c|cc|ccc|ccc|ccc|ccc|ccc|c}
\bottomrule
& \multirow{3}{*}{Separation} & \multirow{3}{*}{Text Query} & \multicolumn{3}{c|}{\multirow{2}{*}{DESED-Test}} & \multicolumn{12}{c|}{WildDESED-Test} & \multirow{2}{*}{Average}\\
& & & & & & \multicolumn{3}{c|}{SNR = 10 dB}  & \multicolumn{3}{c|}{SNR = 5 dB}  & \multicolumn{3}{c|}{SNR = 0 dB}  & \multicolumn{3}{c|}{SNR = -5dB} & \\ 
\cline{4-19}
& & & P1 & P2 & P1+P2 & P1 & P2 & P1+P2 & P1 & P2 & P1+P2 & P1 & P2 & P1+P2 & P1 & P2 & P1+P2 & P1+P2\\
\hline
\#1& \ding{55} & \ding{55} & 0.500 & 0.774 & \textbf{1.274} & 0.329 & 0.591 & \textbf{0.920} & 0.236 & 0.455 & 0.691 & 0.138 & 0.307 & 0.445 & 0.065 & 0.190 & 0.255 & 0.717\\
\hline
\rowcolor[gray]{0.92} 
\#2& \ding{51} & Ground Truth & 0.440 & 0.723 & 1.163 & 0.356 & 0.646 & 1.002 & 0.291 & 0.566 & 0.857 & 0.219 & 0.463 & 0.682 & 0.134 & 0.342 & 0.476 & 0.836\\
\hline
\#3& \ding{51} & All Event Texts & 0.431 & 0.701 & 1.132 & 0.266 & 0.508 & 0.774 & 0.195 & 0.389 & 0.584 & 0.119 & 0.266 & 0.385 & 0.055 & 0.160 & 0.215 & 0.618\\
\hline
\#4& \ding{51} & Ours w/o F.T. & 0.437 & 0.706 & 1.143 & 0.293 & 0.542 & 0.835 & 0.202 & 0.408 & 0.610 & 0.124 & 0.283 & 0.407 & 0.051 & 0.167 & 0.218 & 0.643\\
\#5& \ding{51} & Ours w/o LLM & 0.411 & 0.669 & 1.080 & 0.170 & 0.384 & 0.554 & 0.127 & 0.316 & 0.443 & 0.090 & 0.236 & 0.326 & 0.075 & 0.225 & 0.300 & 0.541\\
\#6& \ding{51} & Our w/o CL& 0.432 & 0.702 & 1.134 & 0.305 & 0.568 & 0.873 & 0.242 & 0.480 & 0.722 & 0.170 & 0.369 & 0.539 & 0.101 & 0.265 & 0.366 & 0.727\\
\#7& \ding{51} & Ours & 0.433 & 0.703 & 1.136 & 0.317 & 0.574 & 0.891 & 0.247 & 0.481 & \textbf{0.728} & 0.177 & 0.370 & \textbf{0.547} & 0.103 & 0.268 & \textbf{0.371} & \textbf{0.735}\\
\toprule

\end{tabular}
}
}
\vspace{-2em}
\label{tab:sed_1}
\end{table*}

\section{Results and Discussions}
In the proposed method, we use LLM-based noise augmentation to fine-tune a SED model and apply a curriculum learning method to enhance the model generalization.
The fine-tuned model is then applied to generate text queries for the LASS model during testing.
To fully explore the effectiveness of the proposed method in noise-robust SED tasks, we present the SED results of different systems in Table~\ref{tab:sed_1}.

\subsection{Using ground truth as text queries}
In \#1, following conventional SED frameworks, we directly input the noisy audio spectrogram into the SED model for predicting frame-wise event results.
While in \#2, we use an LASS model for audio source separation during testing and use weak labels of ground truth as input text queries for the LASS model.
Results show that in noisy environments (i.e., WildDESED-Test), the SED performance is significantly improved by using audio source separation.
By separation and remixing, the interference of noise events is suppressed and the target event components are enhanced, which is beneficial to the detection of target events.
However, in a test environment that is similar to the training environment (i.e., DESED-Test), LASS leads to degraded detection performance.
This is because the DESED-Test contains fewer noise events compared to the WildDESED-Test, resulting in a weaker impact of noise.
Additionally, some target event signal components are distorted during the separation process due to the limited performance of the LASS model. When the benefit from noise suppression is less than the loss of the target signal, the SED performance declines.
Overall, compared to \#1, the average score is improved by 0.119 in terms of P1+P2 in \#2.

\subsection{Using all event texts as text queries}
In \#2, we use weak labels as text queries for the LASS model, which significantly improves the SED performance in noisy environments.
However, in real-world applications, we do not have ground truth labels during testing.
A easiest approach is to use all target event texts as input queries. 
In \#3, we directly input the 10 target event texts into the LASS model, resulting in 10 corresponding sound tracks. 
One obvious drawback of this method is that the LASS model may attempt to isolate a sound event that does not actually exist in the noisy audio clip, and we observe that, in this case, the isolated audio would contain strong noise.
Results in Table~\ref{tab:sed_1} also indicate that in \#3, audio separation lead to a significant decrease in detection performance. 
Compared to \#1, the average score on P1+P2 decreases by 0.099, which suggests that using all event texts as input queries is ineffective.

\subsection{Comparison with Baselines}
Because of the ineffectiveness of using all event texts as text queries, in the proposed system, we propose to use a fine-tuned model for generating text queries.
To explore the impact of fine-tuning, in \#4, we directly use the model trained in stage 1 for predicting clip-wise event results as text queries.
Results show that the average P1+P2 score of \#4 is 0.092 lower than that of \#7, which verifies the superiority of fine-tuning.

\begin{figure}[h]
\centering
\centerline{\includegraphics[width=0.4\textwidth]{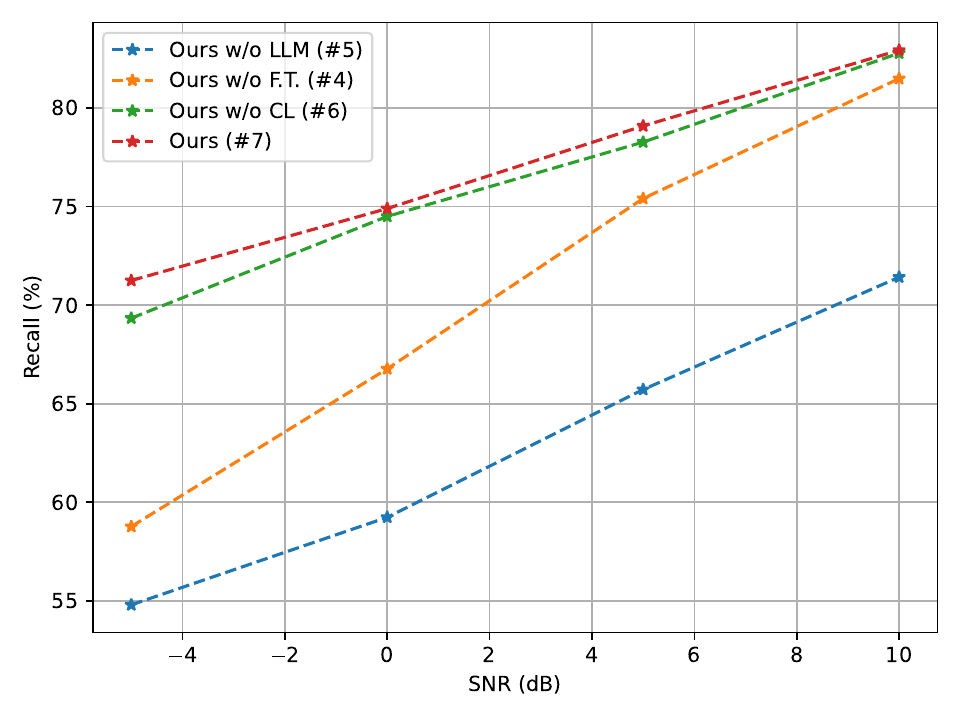}}
\vspace{-1em}
\caption{The macro recall of clip-wise predictions by different systems on WildDESED-Test.}
\label{fig:recall}
\vspace{-1.5em}
\end{figure}

In the proposed method, we use an LLM for picking up noise data during the fine-tuning process.
To illustrate the impact of LLMs, in \#5, we do not use any LLMs but randomly select noise data for fine-tuning.
This approach may result in counter-intuitive combinations. 
{\it For example, the high volume sound of ``wild animals'' appears in an audio clip recorded in a bathroom.}
Results show that the average P1+P2 score of \#5 is significantly lower than that of \#7 and \#4, which indicates that LLMs play an important role in noise augmentation.

To improve the stability of fine-tuning, we use a curriculum learning method to gradually increase the data complexity.
In \#6, the model is fine-tuned without curriculum learning, which means that the SNR is set randomly.
Results show that the detection performance of \#6 is lower than that of \#7, and the average P1+P2 score decreases by 0.008, which verifies the benefits of curriculum learning.

In systems \#4, \#5, \#6, and \#7, we use an SED model to generate clip-wise predictions, which are applied as text queries for the pre-trained LASS model.
In Fig.~2, we present the macro recall of clip-wise predictions predicted by different systems on the WildDESED-Test.
Results show that the proposed system achieves the best recall on all noise levels.
Combining the results in Table.~\ref{tab:sed_1}, we can conclude that higher clip-wise prediction accuracy leads to higher SED scores, and therefore ground truth can help achieve the most superior detection performance among all systems.

\section{Conclusions}
In this paper, we present the effectiveness of LLMs and LASS in improving the performance of SED in changing and noisy real-life applications.
The proposed LLM-based data augmentation method can effectively improve the detection performance in noisy environments, thus providing effective text queries for the LASS model.
Through audio source separation and remixing, the interference from noisy sound events can be effectively suppressed. 

\balance
\bibliographystyle{IEEEtran}
\bibliography{refs}

\begin{thebibliography}{10}
\providecommand{\url}[1]{#1}
\csname url@samestyle\endcsname
\providecommand{\newblock}{\relax}
\providecommand{\bibinfo}[2]{#2}
\providecommand{\BIBentrySTDinterwordspacing}{\spaceskip=0pt\relax}
\providecommand{\BIBentryALTinterwordstretchfactor}{4}
\providecommand{\BIBentryALTinterwordspacing}{\spaceskip=\fontdimen2\font plus
\BIBentryALTinterwordstretchfactor\fontdimen3\font minus \fontdimen4\font\relax}
\providecommand{\BIBforeignlanguage}[2]{{%
\expandafter\ifx\csname l@#1\endcsname\relax
\typeout{** WARNING: IEEEtran.bst: No hyphenation pattern has been}%
\typeout{** loaded for the language `#1'. Using the pattern for}%
\typeout{** the default language instead.}%
\else
\language=\csname l@#1\endcsname
\fi
#2}}
\providecommand{\BIBdecl}{\relax}
\BIBdecl

\bibitem{sed}
A.~Mesaros, T.~Heittola, T.~Virtanen, and M.~D. Plumbley, ``{Sound event detection: a tutorial},'' \emph{IEEE Signal Processing Magazine}, vol.~38, no.~5, pp. 67--83, 2021.

\bibitem{smarthome}
J.~P. Bello, C.~Mydlarz, and J.~Salamon, ``{Sound analysis in smart cities},'' \emph{Springer International Publishing}, pp. 373--397, 2018.

\bibitem{smarthome2}
C.~Debes, A.~Merentitis, S.~Sukhanov, M.~Niessen, N.~Frangiadakis, and A.~Bauer, ``{Monitoring activities of daily living in smart homes: understanding human behavior},'' \emph{IEEE Signal Processing Magazine}, vol.~33, no.~2, pp. 81--94, 2016.

\bibitem{medical}
M.~Vacher, D.~Istrate, L.~Besacier, J.-F. Serignat, and E.~Castelli, ``Sound detection and classification for medical telesurvey,'' in \emph{Proc. International Conference on Biomedical Engineering}, 2004, pp. 395--398.

\bibitem{sed2}
T.~Khandelwal, R.~K. Das, and E.~S. Chng, ``{Sound event detection: a journey through DCASE challenge series},'' \emph{APSIPA Transactions on Signal and Information Processing}, vol.~13, 2024.

\bibitem{crnn4sed}
E.~Cak{\i}r, G.~Parascandolo, T.~Heittola, H.~Huttunen, and T.~Virtanen, ``Convolutional recurrent neural networks for polyphonic sound event detection,'' \emph{IEEE/ACM Transactions on Audio, Speech, and Language Processing}, vol.~25, no.~6, pp. 1291--1303, 2017.

\bibitem{fmsg2}
Y.~Xiao, T.~Khandelwal, and R.~K. Das, ``{FMSG submission for DCASE 2023 challenge task 4 on sound event detection with weak labels and synthetic soundscapes},'' DCASE 2023 Challenge, Tech. Rep., 2023.

\bibitem{fmsg_dcase2024}
Y.~Xiao, H.~Yin, J.~Bai, and R.~K. Das, ``{FMSG-JLESS} submission for {DCASE} 2024 task4 on sound event detection with heterogeneous training dataset and potentially missing labels,'' DCASE 2024 Challenge, Tech. Rep., 2024.

\bibitem{jiangyidi}
{Y. Jiang, R. Tao, W. Huang, Q. Chen and W. Wang}, ``{Unified Audio Event Detection},'' in \emph{Proc. IEEE International Conference on Acoustics, Speech and Signal Processing (ICASSP)}, 2025.

\bibitem{xiao2024mixstyle}
{Y. Xiao, H. Yin, J. Bai, and R. K. Das}, ``Mixstyle based domain generalization for sound event detection with heterogeneous training data,'' \emph{arXiv preprint arXiv:2407.03654}, 2024.

\bibitem{noise1}
M.~Neri, F.~Battisti, A.~Neri, and M.~Carli, ``{Sound event detection for human safety and security in noisy environments},'' \emph{IEEE Access}, vol.~10, pp. 134\,230--134\,240, 2022.

\bibitem{noise2}
T.~Wan, Y.~Zhou, Y.~Ma, and H.~Liu, ``{Noise robust sound event detection using deep learning and audio enhancement},'' in \emph{Proc. IEEE International Symposium on Signal Processing and Information Technology (ISSPIT)}, 2019, pp. 1--5.

\bibitem{noise3}
R.~Serizel, N.~Turpault, A.~Shah, and J.~Salamon, ``{Sound event detection in synthetic domestic environments},'' in \emph{Proc. IEEE International Conference on Acoustics, Speech and Signal Processing (ICASSP)}, 2020, pp. 86--90.

\bibitem{noise4}
Y.~Choi, O.~Atif, J.~Lee, D.~Park, and Y.~Chung, ``{Noise-robust sound-event classification system with texture analysis},'' \emph{Symmetry}, vol.~10, no.~9, p. 402, 2018.

\bibitem{noise5}
I.~McLoughlin, H.~Zhang, Z.~Xie, Y.~Song, and W.~Xiao, ``{Robust sound event classification using deep neural networks},'' \emph{IEEE/ACM Transactions on Audio, Speech, and Language Processing}, vol.~23, no.~3, pp. 540--552, 2015.

\bibitem{gpt4}
J.~Achiam, S.~Adler, S.~Agarwal, L.~Ahmad, I.~Akkaya, F.~L. Aleman, D.~Almeida, J.~Altenschmidt, S.~Altman, S.~Anadkat \emph{et~al.}, ``{GPT}-4 technical report,'' \emph{arXiv preprint arXiv:2303.08774}, 2023.

\bibitem{gpt3}
T.~B. Brown, B.~Mann, N.~Ryder, M.~Subbiah, J.~Kaplan, P.~Dhariwal, A.~Neelakantan, P.~Shyam, G.~Sastry, A.~Askell \emph{et~al.}, ``Language models are few-shot learners,'' in \emph{Proc. International Conference on Neural Information Processing Systems (NIPS)}, 2020, pp. 1877--1901.

\bibitem{llama}
H.~Touvron, T.~Lavril, G.~Izacard, X.~Martinet, M.-A. Lachaux, T.~Lacroix, B.~Rozi{\`e}re, N.~Goyal, E.~Hambro, F.~Azhar \emph{et~al.}, ``{LLaMA}: open and efficient foundation language models,'' \emph{arXiv preprint arXiv:2302.13971}, 2023.

\bibitem{wilddesed}
Y.~Xiao and R.~K. Das, ``{WildDESED}: an {LLM}-powered dataset for wild domestic environment sound event detection system,'' in \emph{Proc. Workshop on Detection and Classification of Acoustic Scenes and Events (DCASE)}, 2024, pp. 196--200.

\bibitem{lass}
X.~Liu, H.~Liu, Q.~Kong, X.~Mei, J.~Zhao, Q.~Huang, M.~D. Plumbley, and W.~Wang, ``Separate what you describe: language-queried audio source separation,'' in \emph{proc. Interspeech}, 2022, pp. 1801--1805.

\bibitem{curriculum1}
Y.~Bengio, J.~Louradour, R.~Collobert, and J.~Weston, ``{Curriculum learning},'' in \emph{Proc. International Conference on Machine Learning (ICML)}, 2009, pp. 41--48.

\bibitem{curriculum2}
S.~Braun, D.~Neil, and S.-C. Liu, ``{A curriculum learning method for improved noise robustness in automatic speech recognition},'' in \emph{Proc. European Signal Processing Conference (EUSIPCO)}, 2017, pp. 548--552.

\bibitem{curriculum3}
D.~Ng, Y.~Xiao, J.~Q. Yip, Z.~Yang, B.~Tian, Q.~Fu, E.~S. Chng, and B.~Ma, ``{Small footprint multi-channel network for keyword spotting with centroid based awareness},'' in \emph{Proc. Interspeech}, 2023, pp. 296--300.

\bibitem{curriculum4}
X.~Wang, Y.~Chen, and W.~Zhu, ``{A survey on curriculum learning},'' \emph{IEEE Transactions on Pattern Analysis and Machine Intelligence}, vol.~44, no.~9, pp. 4555--4576, 2021.

\bibitem{curriculum5}
A.~Pentina, V.~Sharmanska, and C.~H. Lampert, ``{Curriculum learning of multiple tasks},'' in \emph{Proc. IEEE Conference on Computer Vision and Pattern Recognition (CVPR)}, 2015, pp. 5492--5500.

\bibitem{dcase2024T4}
S.~Cornell, J.~Ebbers, C.~Douwes, I.~Mart{\'\i}n-Morat{\'o}, M.~Harju, A.~Mesaros, and R.~Serizel, ``Dcase 2024 task 4: Sound event detection with heterogeneous data and missing labels,'' \emph{arXiv preprint arXiv:2406.08056}, 2024.

\bibitem{audioset}
J.~F. Gemmeke, D.~P.~W. Ellis, D.~Freedman, A.~Jansen, W.~Lawrence, R.~C. Moore, M.~Plakal, and M.~Ritter, ``{Audio set: an ontology and human-labeled dataset for audio events},'' in \emph{Proc. IEEE International Conference on Acoustics, Speech and Signal Processing (ICASSP)}, 2017, pp. 776--780.

\bibitem{psds1}
{\c{C}}.~Bilen, G.~Ferroni, F.~Tuveri, J.~Azcarreta, and S.~Krstulovi{\'c}, ``A framework for the robust evaluation of sound event detection,'' in \emph{Proc. IEEE International Conference on Acoustics, Speech and Signal Processing (ICASSP)}, 2020, pp. 61--65.

\bibitem{psds2}
J.~Ebbers, R.~Haeb-Umbach, and R.~Serizel, ``Threshold independent evaluation of sound event detection scores,'' in \emph{Proc. IEEE International Conference on Acoustics, Speech and Signal Processing (ICASSP)}, 2022, pp. 1021--1025.

\bibitem{beats}
S.~Chen, Y.~Wu, C.~Wang, S.~Liu, D.~Tompkins, Z.~Chen, W.~Che, X.~Yu, and F.~Wei, ``Beats: audio pre-training with acoustic tokenizers,'' in \emph{Proc. International Conference on Machine Learning (ICML)}, 2023, pp. 5178--5193.

\bibitem{audiosep-dp}
H.~Yin, J.~Bai, Y.~Xiao, H.~Wang, S.~Zheng, Y.~Chen, R.~K. Das, C.~Deng, and J.~Chen, ``Exploring text-queried sound event detection with audio source separation,'' in \emph{Proc. IEEE International Conference on Acoustics, Speech and Signal Processing (ICASSP)}, 2025.

\bibitem{dprnn}
Y.~Luo, Z.~Chen, and T.~Yoshioka, ``Dual-path {RNN}: efficient long sequence modeling for time-domain single-channel speech separation,'' in \emph{Proc. IEEE International Conference on Acoustics, Speech and Signal Processing (ICASSP)}, 2020, pp. 46--50.

\bibitem{clap}
Y.~Wu, K.~Chen, T.~Zhang, Y.~Hui, T.~Berg-Kirkpatrick, and S.~Dubnov, ``Large-scale contrastive language-audio pretraining with feature fusion and keyword-to-caption augmentation,'' in \emph{Proc. IEEE International Conference on Acoustics, Speech and Signal Processing (ICASSP)}, 2023, pp. 1--5.

\bibitem{mean-teacher}
J.~Deng, W.~Li, Y.~Chen, and L.~Duan, ``Unbiased mean teacher for cross-domain object detection,'' in \emph{Proc. IEEE/CVF Conference on Computer Vision and Pattern Recognition (CVPR)}, 2021, pp. 4091--4101.

\end{thebibliography}

\end{document}